\documentclass{aa} 
\usepackage{txfonts}
\usepackage{hyperref}
\usepackage{threeparttable}
\begin{document}

   \title{Testing massive star evolution, star-formation history, and feedback at low metallicity}

   \subtitle{Photometric analysis of OB stars in the SMC Wing \thanks{Based in part on observations made with the Southern African Large Telescope (SALT)}}

   \author{Leah M. Fulmer
          \inst{1},\inst{2}
          John S. Gallagher, III
          \inst{3},
          Wolf-Rainer Hamann
          \inst{4},
          Lida M. Oskinova
          \inst{4},
          \and
          Varsha Ramachandran\inst{4}}

   \institute{University of Washington, Physics-Astronomy Bldg. 3910 15th Ave NE Seattle, WA 98195, USA; lfulmer@uw.edu
              \and
              ARCS Foundation Fellow
              \and
              Department of Astronomy, University of Wisconsin - Madison, WI 53706, USA
              \and
             Institute of Physics and Astronomy, University of Potsdam, Potsdam 14457, Germany\\
             }

   \date{Received September 25, 2018; Accepted December 12, 2019}
 
  \abstract
   {The supergiant ionized shell SMC-SGS~1 (DEM 167), which is located in the outer Wing of the Small Magellanic Cloud (SMC), resembles structures that originate from an energetic star-formation event and later stimulate star formation as they expand into the ambient medium. However, stellar populations within and surrounding SMC-SGS~1 tell a different story.}
   {We present a photometric study of the stellar population encompassed by SMC-SGS~1 in order to trace the history of such a large structure and its potential influence on star formation within the low-density, low-metallicity environment of the SMC.}
   {For a stellar population that is physically associated with SMC-SGS~1, we combined near-ultraviolet (NUV) photometry from the Galaxy Evolution Explorer (GALEX) with archival optical (V-band) photometry from the ESO Danish 1.54m Telescope. Given their colors and luminosities, we estimated stellar ages and masses by matching observed photometry to theoretical stellar isochrone models.}
   {We find that the investigated region supports an active, extended star-formation event spanning $\sim$ 25 - 40 Myr ago, as well as continued star formation into the present. Using a standard initial mass function (IMF), we infer a lower bound on the stellar mass from this period of $\sim 3 \times 10^4~M_{\odot}$, corresponding to a star-formation intensity of $\sim$ 6 $\times$ 10$^{-3}$~M$_{\odot}$~kpc$^{-2}$~yr$^{-1}$.}
   {The spatial and temporal distributions of young stars encompassed by SMC-SGS~1 imply a slow, consistent progression of star formation over millions of years. Ongoing star formation, both along the edge and interior to SMC-SGS~1, suggests a combined stimulated and stochastic mode of star formation within the SMC Wing. We note that a slow expansion of the shell within this low-density environment may preserve molecular clouds within the volume of the shell, leaving them to form stars even after nearby stellar feedback expels local gas and dust.}

\authorrunning{Fulmer et. al.}
\titlerunning{Star formation and feedback at low metallicity}

   \maketitle

\section{Introduction}\label{sec:introduction}
\begin{figure*}[!ht]
  \centering
  \includegraphics[width=0.454\textwidth]{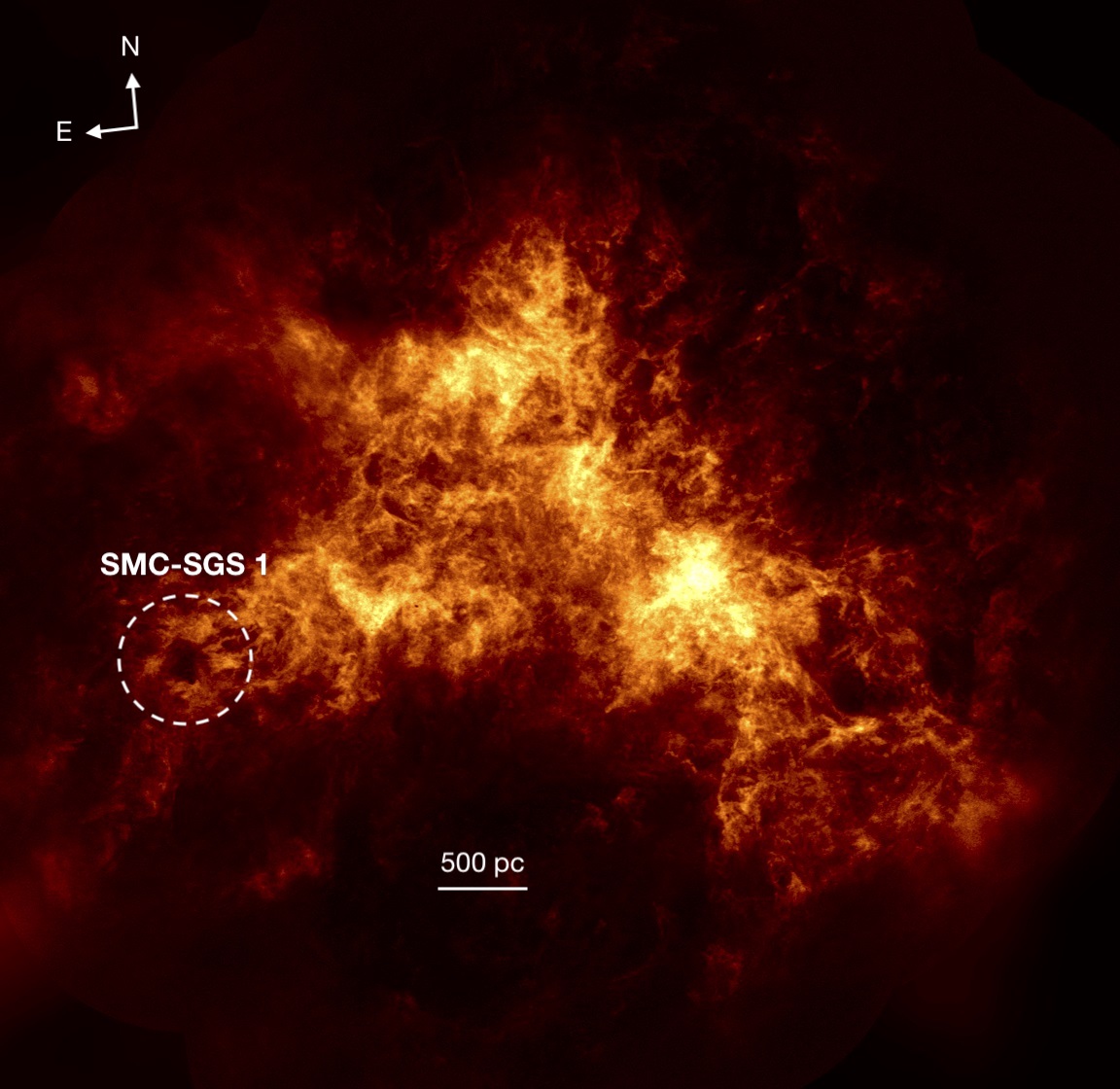}
  \includegraphics[width=0.527\textwidth]{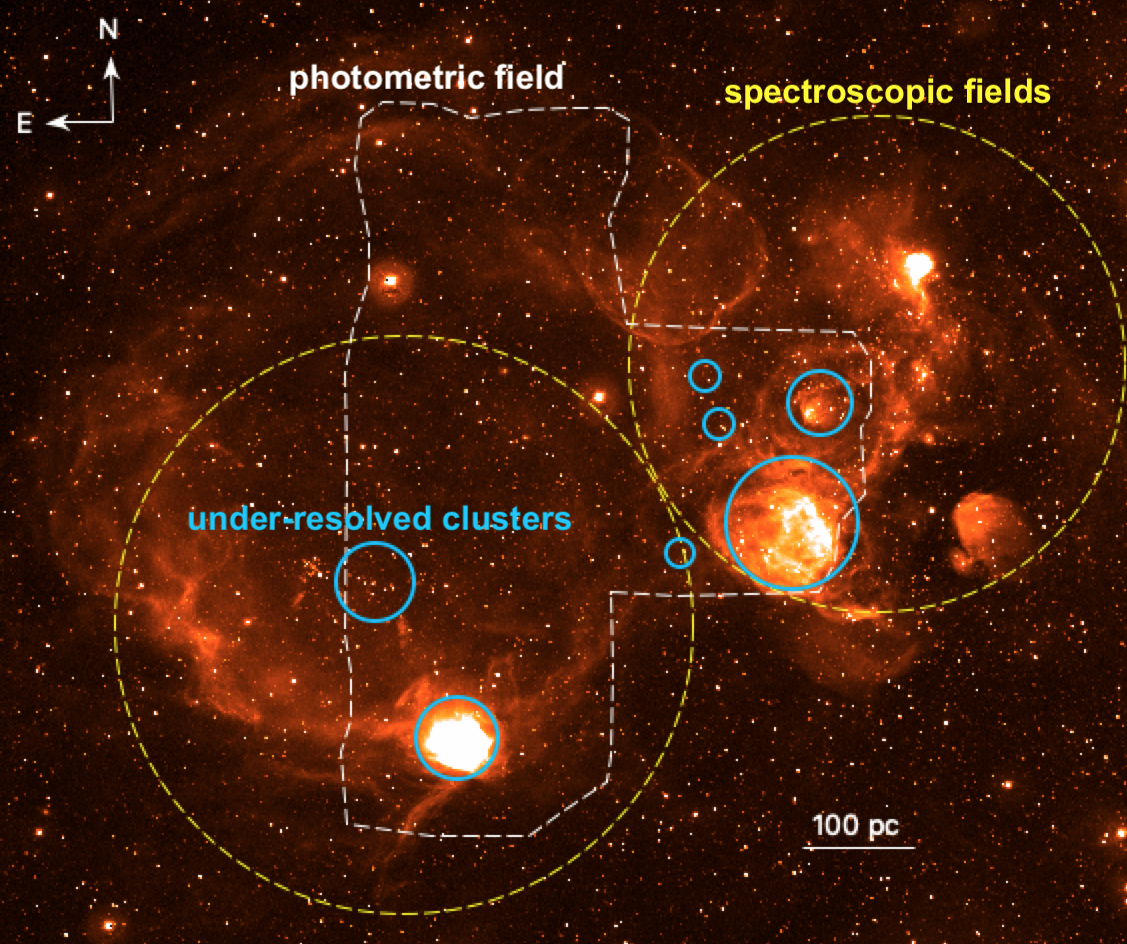}
  \caption{(a) Australian Square Kilometer Array Pathfinder (ASKAP) H\,{\sc i} image of the SMC Wing, with the supergiant shell SMC-SGS~1 at its southeastern tip \citep{McClureGriffiths18}. (b) Narrow-band H$\alpha$ image of SMC-SGS~1 from the Magellanic Cloud Emission-Line Survey (MCELS, \citet{2005AAS...20713203W}) with regions indicating the photometric field (white), \citet{2019AA...625A.104R} spectroscopic fields (yellow), and under-resolved clusters (blue).}
  \label{fig:smc_investigated_region}
    \end{figure*}

\textit{Overview.} 
The Wing of the Small Magellanic Cloud (SMC, Figure~\ref{fig:smc_investigated_region}a) hosts ongoing, active star formation within an environment of diffuse, low-metallicity interstellar matter (ISM). Given its proximity (distance modulus $\approx$ 18.7 mag) and relative isolation, the SMC Wing is often considered to be an ideal laboratory through which to study star formation under such conditions. Investigations within this region seek to uncover the mechanism behind massive star formation within the greater narrative of star formation in low-density, low-metallicity environments \citep[][]{1964MNRAS.127..429W}.

\textit{SMC-SGS~1.} 
The SMC Wing hosts the supergiant ionized shell SMC-SGS~1, which is characterized as a large, low-brightness structure that hosts the Wing's most notable star cluster NGC~602 on its southern edge \citep{1976MmRAS..81...89D}. Despite its size (radius $\approx$ 300~pc), SMC-SGS~1 maintains a relatively symmetric circular structure in ionized gas \citep{1980MNRAS.192..365M}. The presence of such a large, well-defined structure implies that SMC-SGS~1 formed after an energetic star formation event isotropically expelled local gas and dust, thus creating and ionizing our gaseous shell. \citet{1995AJ....109..594K} estimated the H$\alpha$ luminosity of SMC-SGS~1, and concluded that the structure requires only a few O stars to account for its ionization. This region contains a well-known WO binary star \citep{2016AA...591A..22S}, which \citet{2019AA...625A.104R} showed to provide the large majority of ionizing radiation. Optical and H\,{\sc i} radial velocity measurements of SMC-SGS~1 place the shell at V$_{helio} = $175~km~s$^{-1}$ with an expansion velocity of $\approx$ 10~km~s$^{-1}$ based on the H\,{\sc i} maps from \citet{1997MNRAS.289..225S} and \citet{1999MNRAS.302..417S}. The NGC~602 H\,{\sc ii} complex located along the southern edge of SMC-SGS~1 demonstrates V$_{helio}$ = $179 \ \pm$ 5~km~s$^{-1}$, and it is thus physically associated with the larger structure \citep{2008PASP..120..972N}.

\textit{Local interstellar medium.} 
As emphasized by \citet{2008PASP..120..972N}, the interstellar medium surrounding SMC-SGS~1 consists of multiple shell structures with various velocity components. H\,{\sc i} features are found over the velocity range of $\sim$140 - 200~km~s$^{-1}$, which is similar to the range found for ionized gas by \citet{1995RMxAC...3..255A}. In particular, various structures along the western edge of SMC-SGS~1 (Figure~\ref{fig:smc_investigated_region}b) are associated with H\,{\sc i} gas of V$_{helio}$ $\approx$ 150-170~km~s$^{-1}$. This type of environment likely results from a combination of multiple tidal features that are observed in projection, as well as interactions between expanding shells. Existing X-ray observations do not reveal the presence of a hot medium filling the shell \citep{2013ApJ...765...73O}, and thus further support the suggestion that this is a relatively old feature. However, despite the apparent complexity in this region, \citet{2010ApJ...708.1204B}, \citet{2017ApJ...845...53N}, and \citet{2019arXiv191104370S} find relatively low levels of turbulence throughout the SMC Wing, which is consistent with conditions that produce large-scale galactic tidal features.

\textit{NGC~602.} 
For over half a century, the SMC Wing has been known to contain young stars despite a distinctive lack of dense interstellar gas and dust \citep[e.g.,][]{1958MNRAS.118..172L, 1964MNRAS.127..429W, 1971AA....10....1W, 1978MNRAS.184..631N, 1985PASP...97..530H, 1986AA...161..408P}. Within the last decade, a series of detailed investigations outlined an evolutionary history for NGC~602 and uncovered two distinct episodes of star formation. The latter episode began $\sim$2 Myr ago and continues into the present \citep{2009AJ....137.3668C, 2013ApJ...775...68D}. The former episode occurred $\sim$30 Myr ago, but it extended over a much wider region of the SMC Wing than a single star cluster. Given the location of NGC~602 at the edge of SMC-SGS~1, the cluster may have formed as a result of sequential star formation at the expanding shell edge. This process of sequential star formation would have impacted the SMC Wing at large, and may have been reinforced by interactions with H\,{\sc i} shells within the SMC Wing \citep{1997MNRAS.289..225S, 2008PASP..120..972N}.

\textit{Larger stellar population.} 
\citet{2019AA...625A.104R} obtain and analyze high-resolution spectra for 320 massive stars encompassed by SMC-SGS~1 in order to investigate the effect of a low-metallicity environment on stellar feedback, star formation, and late-stage evolution. Although the \citet{2019AA...625A.104R} investigation specifically targets massive stars, their spectroscopic sample overlaps with our photometric survey (Figure~\ref{fig:smc_investigated_region}b). They observe a stochastic mode of massive star formation which, when combined with low stellar feedback and metal-poor stellar populations, supports extended star-formation events over long timescales. Their spectroscopic stellar ages reveal one such event spanning $\sim$ 30 to 40 Myr ago. Their analysis of stellar feedback from both supernovae and the WO star in this region illustrate that these alone would provide sufficient mechanical energy to sustain large gaseous structures, such as SMC-SGS~1.

\textit{Comparison to dwarf irregular galaxies.} 
The Galaxy Evolution Explorer (GALEX) ultraviolet survey uncovered wide-spread star formation in dwarf systems \citep{2009ApJ...706..599L} and the low-density outskirts of galaxies \citep[e.g,][]{2010AJ....140.1194B,2010AJ....139..447H}. Observations of irregular galaxies by \citet{2016AJ....151..136H} show that young stars can be found in regions where H\,{\sc i} surface densities are as low as $\leq$ 1~M$_{\odot}$~pc$^{-2}$. The region containing SMC-SGS~1 has an average H\,{\sc i} column density of $\leq 5 \times 10^{21}$~atoms~cm$^{-2}$ \citep{1999MNRAS.302..417S}, which corresponds to an H\,{\sc i} surface density of $ \leq$ 3~M$_{\odot}$~pc$^{-2}$ when one adopts the conversion by \citet{2016AJ....151..136H}. Therefore, the SMC may be among a population of gas-rich dwarf irregular galaxies that demonstrate star formation throughout their outer disks \citep[e.g.,][]{ 2018AJ....156...21H}, as well as in tidal debris \citep[e.g.,][]{2008AJ....135..548D}.

\textit{This investigation.} 
Building upon the spectroscopic investigation by \cite{2019AA...625A.104R}, we present a wide-spread near-ultraviolet (NUV) photometric stellar population study of the SMC Wing. Our investigated region covers an area roughly 40 times that of studies focusing primarily on NGC~602. Due to this coverage, we are able to investigate star formation both within the volume and along the edge of SMC-SGS~1 and thus clarify the influence of the shell in this region's star-formation history. Our investigated region intersects that of \cite{2019AA...625A.104R}, and we take their findings as largely representative of young stars within our population. That being said, we recognize that some discrepancies may occur between our results because neither the investigated regions nor the targeted populations are exactly congruent (the spectroscopic study specifically targets bright, massive stars, whereas our photometric survey additionally includes relatively faint sources). Our combination of photometric, spectroscopic, theoretical, and spatial analyses, offers a multi-faceted illustration of star-formation patterns and their connection to gaseous structures in the SMC Wing. Section \ref{sec:observations} describes the observations and models demonstrated in this work. Section \ref{sec:analysis} describes our various analyses \footnote{\href{https://github.com/lfulmer/star-formation-smc-wing}{https://github.com/lfulmer/star-formation-smc-wing}}. Section \ref{sec:results} presents the results from these analyses, as well as a discussion of their physical implications. Section \ref{sec:conclusions} summarizes our conclusions.

\section{Observations and models}\label{sec:observations}

\begin{table}[!ht]
\centering
\begin{threeparttable}
  \caption{Observational data sets in our photometric survey}
  \label{tab:observational_properties}
  \begin{tabular}{lll}
    \hline
\noalign{\smallskip}
Property & (NUV)$_\circ$ & (V)$_\circ$ \\
        \hline
        \noalign{\smallskip}
        Telescope & GALEX & Danish 1.54-m \\
        \noalign{\smallskip}
        Mean wavelength [$\mbox{\AA}$] & 2320 & 5500 \\
        \noalign{\smallskip}
        Resolution [arcsec] & $\sim$ 5 & $\sim$ 1.5 \\
        \noalign{\smallskip}
        Source extraction & IRAF {\tt DAOPHOT} & (1)\\
        \hline
        \noalign{\smallskip}
        \end{tabular}
        \begin{tablenotes}
        \small
        \item References. (1) \citet{2001PhDT.........3B}.
        \end{tablenotes}
\end{threeparttable}
\end{table}

\textit{Overview.} 
In order to highlight recent star formation within the SMC Wing, we obtained both NUV and optical (V-band) photometry for this investigation. We extracted NUV photometry from a mosaic of archival GALEX images including SMC-SGS~1. Our V-band photometry and uncertainties were adopted from the \citet{2001PhDT.........3B} survey of the Magellenic Clouds, observed with the ESO Danish 1.54m Telescope.

\textit{NUV image reduction.} 
Our wide-field GALEX mosaic of the SMC Wing was constructed from co-added tiles available from the Mikulski Archive for Space Telescopes (MAST) GALEX science archive\footnote{\href{http://galex.stsci.edu/}{http://galex.stsci.edu/}}. After splitting the tiles by bandpass into a set of NUV and FUV exposures, a mosaic of NUV exposures was created using {\tt swarp} \citep{2002ASPC..281..228B}. To account for different integration times and the spatially-dependent response function, archival high-resolution response function images were used as weights and the weighted mean was calculated when co-adding individual images. As all GALEX tiles are already photometrically calibrated, no additional flux corrections were applied to individual frames, and no additional background subtraction was required.

\textit{NUV source extraction.} 
The GALEX archival NUV catalog does not provide a sufficient population of photometered stars from which to proceed with a meaningful study of the region containing SMC-SGS~1; therefore, we chose to photometer stars with IRAF {\tt DAOPHOT}. Given the variable crowding within our investigated region, as well as GALEX image edge effects, we adopt a conservative error of $\pm$0.1 dex for our GALEX NUV stellar photometry, which is roughly twice the uncertainty quoted by \citet{2007ApJS..173..682M} for the GALEX All-sky UV Survey. Within fields of significant blending, {\tt DAOPHOT} could not resolve individual stars. This became a particular issue in regions of high stellar surface density, including among the star cluster NGC~602. Therefore, our analysis includes an evaluation of the luminosity among both resolved, ``photometered'' stars and ``under-resolved'' clusters in the investigated region (Figure~\ref{fig:smc_investigated_region}b). 

\begin{table}[!ht]
  \centering
  \begin{threeparttable}
  \caption{Photometric corrections for resolved stars}
  \label{tab:photometric_corrections}
  \begin{tabular}{lll}
    \hline
\noalign{\smallskip}
Correction type [AB mag] & (NUV)$_\circ$ & (V)$_\circ$ \\
        \hline
        \noalign{\smallskip}
        Distance modulus  & 18.7  & 18.7\\
        \noalign{\smallskip}
        Galactic extinction  & -0.42  & -0.16\\
        \noalign{\smallskip}
        A$_\lambda$ / E$_{B-V}$ (1)  & 8.08  & 2.93 \\
        \noalign{\smallskip}
        GALEX catalog comparison & 0.6  & None\\
        \hline
        \noalign{\smallskip}
        \end{tabular}
        \begin{tablenotes}
        \small
        \item References. (1) \citep{2011ApSS.335...51B}.
        \end{tablenotes}
        \end{threeparttable}
\end{table}

\textit{Photometric corrections.} 
To determine our photometric sample, we first matched the NUV and V-band stellar catalogs with TOPCAT \citep{2005ASPC..347...29T} according to their J2000 coordinates. We then applied photometric corrections to our data set, adjusting for Galactic extinction, local SMC extinction, and systematic discrepancies between the {\tt DAOPHOT} and GALEX catalog photometries (Table~\ref{tab:photometric_corrections}). To correct for Galactic extinction, we assumed a foreground screen correction with $E(B-V)$ = 0.052. For local extinction within the SMC, we divided the investigated region into four individual subregions: one located at the center of SMC-SGS~1 and three at its northern, western, and southern edges. We applied a foreground screen correction to the subregions individually, assuming minimal extinction toward the center of SMC-SGS~1 ( $E(B-V)$ = 0 ) and adjusting edge extinction corrections to match it in color-magnitude space. We adopted the \citet{2011ApSS.335...51B} selective extinction for the SMC and found extinction values of $E(B-V)$ = 0.04, 0.08, and 0 for the northern, western and southern subregions, respectively. We removed stars from our population with proper motion < 4.0 mas/year \citep{2018AA...616A...1G} to eliminate foreground stars. Finally, we adopted an apparent magnitude detection limit of NUV $\approx$ 19.5, which is consistent with other GALEX NUV photometric studies of the SMC \citep{Simons14}. Following these corrections, our photometric sample consisted of $\sim$1100 sources.

\textit{Spectroscopic observations.} 
Spectroscopy provided radial velocity information for the gaseous and stellar components of the SMC Wing and augmented our photometric analysis where there were common sources with \citet{2019AA...625A.104R}. H\,{\sc i} velocities were measured from the \citet{1999MNRAS.302..417S} H\,{\sc i} data cube, which we analyzed over our investigated region (Figure~\ref{fig:smc_investigated_region}b, white). H$\alpha$ observations were obtained with the Southern African Large Telescope (SALT) on November 1, 2016 using the Robert Stobie Spectrograph with slit dimensions 8' $\times$ 1.25". The slit was oriented with its center at J2000 $\alpha$ = 01:32:42.8, $\delta$ = -73:27:10 and a position angle of 87$^\circ$, extending from the western edge of SMC-SGS~1 inward. A 2300 mm$^{-1}$ grating yielded a resolution of 4400 ($\sigma =$ 35~km~s$^{-1}$) in the H$\alpha$ spectral region. Four exposures of this spectra were taken for a total integration time of 4400~s. Stellar radial velocities and classifications were adopted from \citet{2019AA...625A.104R}. The combined photometric and spectroscopic sample contains $\sim$100 stars or $\sim$10\% of the photometered population.

\begin{figure}[!ht]
  \includegraphics[width=\columnwidth]{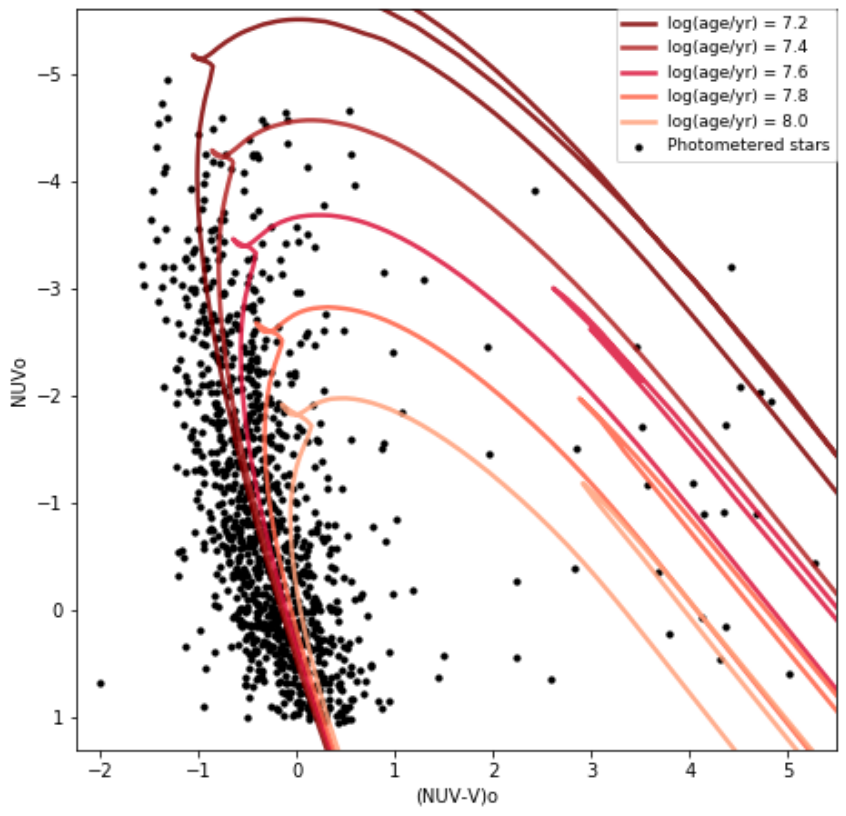}
    \caption{Points: NUV$_\circ$ vs. (NUV-V)$_\circ$ color-magnitude diagram (CMD) for $\sim$ 1100 stars in the SMC Wing. Tracks: Isochrone models for stars of the same age and metallicity, with various masses (Padova).}
    \label{fig:cmd_isochrones}
\end{figure}

\begin{table}[!ht]
  \centering
\begin{threeparttable}
  \caption{Isochrone model input parameters}
  \label{tab:isochrone_metadata}
  \begin{tabular}{p{2.3cm}p{5.8cm}}
    \hline
\noalign{\smallskip}
Input parameter & Value \\
        \hline\noalign{\smallskip}
        Model reference & (1) \\\noalign{\smallskip}
        Evolutionary tracks & PARSEC v1.2S (2), COLIBRI PR16 (3) \\\noalign{\smallskip}
        Photometric system & GALEX FUV, NUV, Johnson's UBV (4), Vega magnitudes converted to AB mags \\\noalign{\smallskip}
        IMF & lognormal (5) \\
        \noalign{\smallskip}
        Ages & log(age/year) = [7.2,7.4,7.6,7.8,8.0] \\
        \noalign{\smallskip}
        Metallicity & 0.004 \\
        \hline
        \noalign{\smallskip}
        \end{tabular}
        \begin{tablenotes}
        \small
        \item References. (1) \citet{2017ApJ...835...77M}.; (2) \citep{2012MNRAS.427..127B}; (3) \citep{2013MNRAS.434..488M}; (4) \citep{2006AJ....131.1184M}; (5) \citep{2001ApJ...554.1274C}.
        \end{tablenotes}
        \end{threeparttable}
\end{table}

\textit{Theoretical models.} 
Isochrones describe the theoretical colors and luminosities for a series of stars with the same age and various initial masses. In this study, isochrones were adopted from the University of Padova\footnote{\href{http://stev.oapd.inaf.it/cgi-bin/cmd}{http://stev.oapd.inaf.it/cgi-bin/cmd}} \citep{2017ApJ...835...77M}. When superimposed on our NUV$_\circ$ vs. (NUV-V)$_\circ$ color-magnitude diagram (CMD), they effectively quantify relative age patterns within our sample without ever being fit to our data directly. Table~\ref{tab:isochrone_metadata} describes input parameters for the isochrone models used in this investigation, and Figure~\ref{fig:cmd_isochrones} (tracks) exhibits them superimposed on the NUV$_\circ$ vs. (NUV-V)$_\circ$ CMD.

\section{Analysis}\label{sec:analysis}

\textit{Stellar luminosity.} 
Because source extraction with {\tt DAOPHOT} could not resolve all of the stars within our investigated region, we pursued independent analyses of the entire investigated region, the resolved stellar population, and the under-resolved star clusters. We conducted an {\tt Apphot} analysis of the entire photometric region (Figure~\ref{fig:smc_investigated_region}b, white), for which we applied a background correction based on quiescent regions to the east and south. We consider this analysis to account for 100\% of the stellar luminosity in our investigation. For stars resolved and photometered by {\tt DAOPHOT}, Figure~\ref{fig:cmd_isochrones} (points) illustrates the (NUV)$_\circ$ vs. (NUV-V)$_\circ$ CMD, which reveals relative ages and evolutionary trends among our stellar population. Resolved stellar luminosities were directly converted from {\tt DAOPHOT} (NUV)$_\circ$ AB magnitudes to counts per second. The sum of these values recovered $\sim$70\% of the stellar luminosity within our investigated region as compared to the {\tt Apphot} analysis of the entire region. We located seven star clusters that were under-resolved in our source extraction (Figure~\ref{fig:smc_investigated_region}b, blue). These clusters were selected by visual inspection; we examined regions exhibiting extended luminosity contours and identified which of the regions contained an apparent underestimate of resolved stars. We summed the resulting luminosity from each of these seven clusters and subtracted the luminosity from stars that had been resolved within them. We found that the clusters recovered the remaining $\sim$30\% of the luminosity in our investigated region.

\begin{figure}[!ht]
\includegraphics[width=\columnwidth]{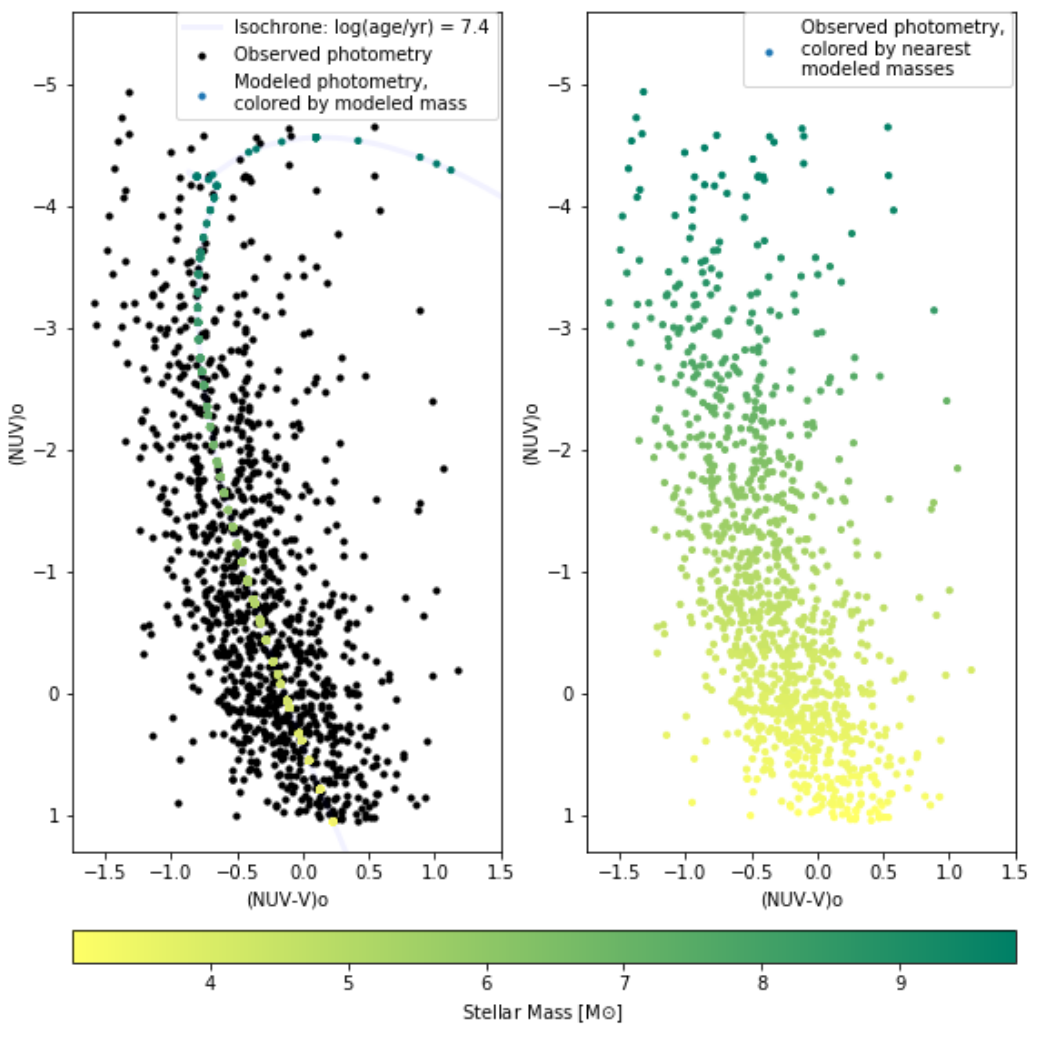}
    \caption{(a) Black: (NUV)$_\circ$ vs. (NUV-V)$_\circ$ CMD for our photometered main sequence stars. Green: Theoretical NUV vs. NUV-V photometry from isochrone models, colored by their associated theoretical masses. (b) Same photometry, colored by the stellar mass associated with each object's closest theoretical NUV magnitude.}
    \label{fig:mass_definition}
\end{figure}

\textit{Stellar mass definition.} 
We estimated stellar masses for our photometered stars by matching the observed NUV$_\circ$ photometry to theoretical NUV photometry along an isochrone of log (age/year) = 7.4 (Figure~\ref{fig:mass_definition}a). Because our population included stars of various ages, we recognized that no single isochrone would provide a perfect representation of stars within this region. However, to estimate the mass of young stars within our investigated region, this isochrone appropriately represents the youngest age along which we observe a photometric departure from the main sequence (Figure~\ref{fig:cmd_isochrones}). We matched our observed NUV$_\circ$ observations to the closest theoretical photometric values along the chosen isochrone. Then, we used the stellar mass estimate associated with each theoretical magnitude to determine an estimated mass for each star (Figure~\ref{fig:mass_definition}b). We denote the resulting masses as ``observed'' stellar masses, because they are fundamentally based on a match to the observed stellar photometry. 

\textit{Stellar mass calculation.} 
To calculate the mass of young stars in our investigated region, we adopted a Salpeter-like relationship for the upper initial mass function (IMF) \citep{2001ApJ...554.1274C}. We calculated the ratio of observed stellar mass to theoretical stellar mass by integrating stellar masses proportional to the IMF for all observed and theoretical masses, respectively. Within the integration, we defined the maximum observed stellar mass as the largest mass within the isochrone match analysis and the maximum theoretical stellar mass as the largest spectroscopically-derived stellar mass from among the combined photometric and spectroscopic sample. With the ratio of observed to theoretical stellar mass, we converted our observed stellar mass from the isochrone match analysis to a theoretical mass for young stars in the investigated region (Table~\ref{tab:mass_calculation}). We consider this value to be a lower bound on the stellar mass encompassed by SMC-SGS~1 because young, massive stars are more likely to be found in high-density clusters, which {\tt DAOPHOT} systematically under-resolved.

\begin{figure*}[!ht]
  \includegraphics[width=0.54\textwidth]{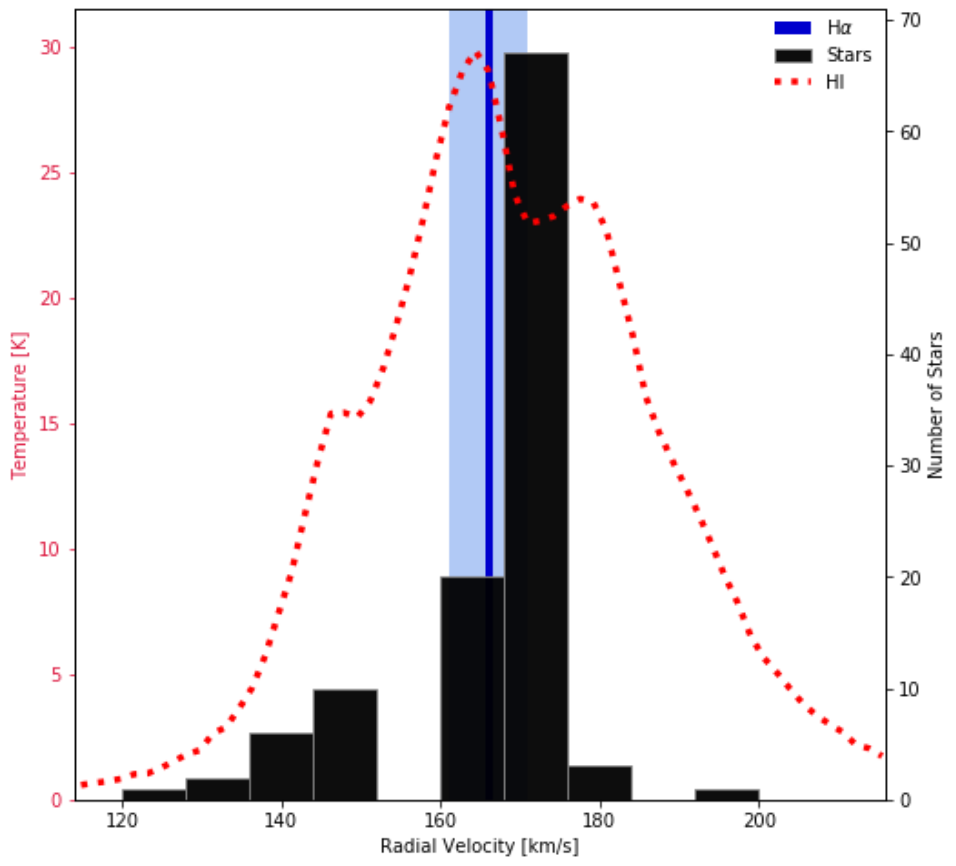}
  \includegraphics[width=0.455\textwidth]{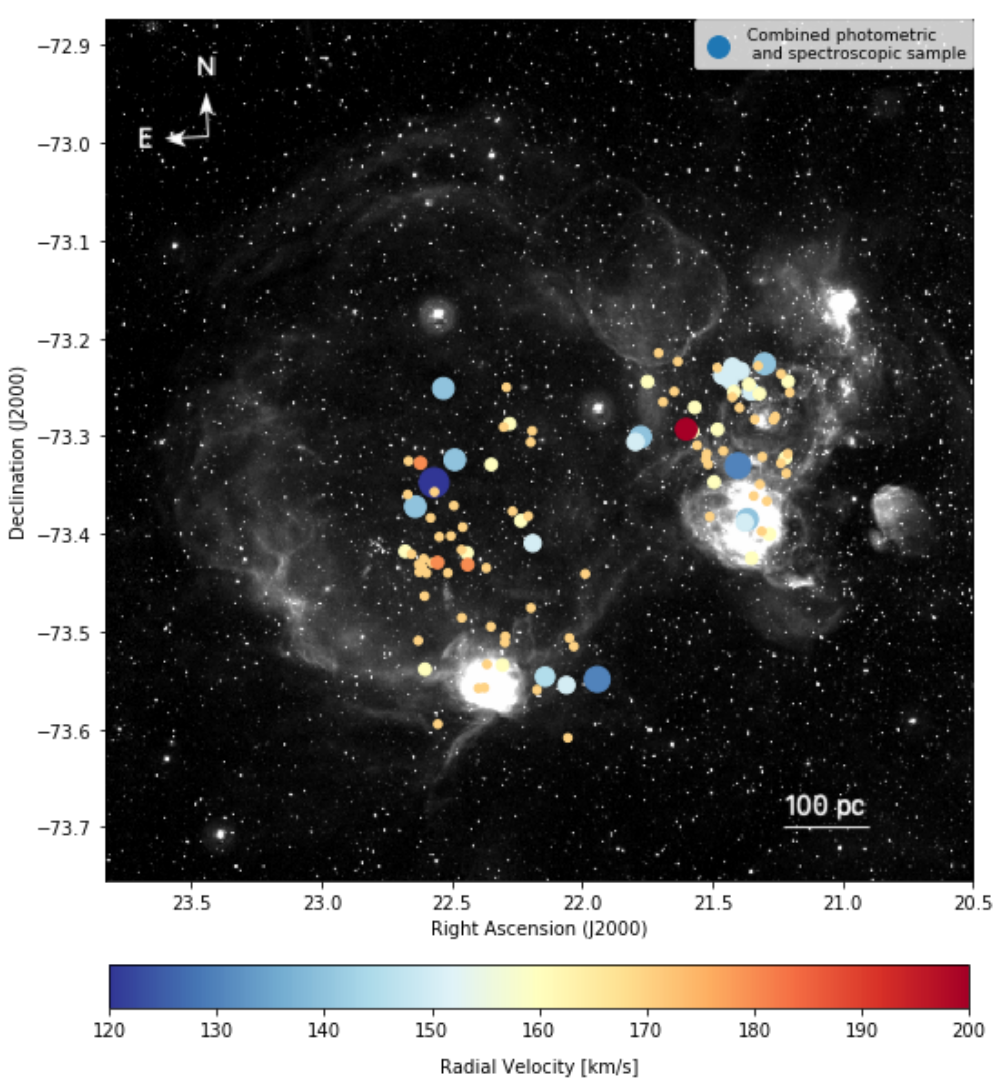}
    \caption{(a) Heliocentric radial velocities for our investigated region. Blue: A single radial velocity measurement and the associated uncertainty for H$\alpha$, observed with the Southern African Large Telescope (SALT). Red: Temperature vs. radial velocity for H\,{\sc i} gas within the investigated region, observed with the Australia Telescope Compact Array (ATCA). Black: Histogram of stellar radial velocities from \citet{2019AA...625A.104R}, covering a population of $\sim$100 photometered stars, or $\sim$10\% of the photometered population. (b) MCELS narrow-band H$\alpha$ image of the SMC Wing highlighting the supergiant shell SMC-SGS~1. Points indicate the locations of stars among the combined photometric and spectroscopic sample, with marker color representing radial velocity and marker size representing difference from the stellar median radial velocity of 170 km~$s^{-1}$.}
    \label{fig:velocities}
\end{figure*}

\textit{Gaseous and stellar radial velocities.} 
We looked for synergies between the gaseous and stellar components of our investigated region to ensure that there was clear evidence for their physical association. Figure~\ref{fig:velocities}a shows heliocentric radial velocities for H$\alpha$ gas (blue), H\,{\sc i} gas (red), and stars (black) within the combined photometric and spectroscopic sample.

\textit{Stellar classification.} 
Spectroscopic studies provide more precise stellar classifications than photometric observations, and as such are invaluable complements to any photometric stellar survey. We adopted stellar classifications from \citet{2019AA...625A.104R} where there were common sources, leading to a combined photometric and spectroscopic sample of $\sim$ 100 stars. Figure~\ref{fig:photometric_spectroscopic_spatial}a illustrates the NUV$_\circ$ vs. (NUV-V)$_\circ$ CMD with all photometered stars shown in gray and spectroscopically-classified stars indicated with various markers.

\textit{Spatial analyses.} 
To understand the relationship between the star-formation history of our population and the surrounding gaseous environment, we performed various analyses combining photometric, spectroscopic, and spatial information. Figure~\ref{fig:velocities}b offers a spatial analysis of the stellar radial velocity histogram. Within this figure, stellar velocities are represented with various colors, and the difference between an individual star's radial velocity and the population median radial velocity is shown through marker size. Figure~\ref{fig:photometric_spectroscopic_spatial}b illustrates a spatial analysis with stellar classifications represented with marker color and shape. Figure~\ref{fig:photometric_spatial} demonstrates a combined photometric and spatial analysis, in which marker colors and sizes correlate with stellar (NUV)$_\circ$ magnitude. This figure illustrates the projected distribution of stellar luminosities throughout our investigated region.

\section{Results and associated discussion}\label{sec:results}

\subsection{The stellar population and its star-formation history}

\textit{Overview.} 
Photometered stars with relatively blue (NUV-V)$_\circ$ colors and bright NUV$_\circ$ magnitudes reveal significant populations of young, massive stars encompassed by SMC-SGS~1 (Figure~\ref{fig:cmd_isochrones}). Simultaneously, stars to the right of the main sequence plume give evidence for evolved stars of various ages. The broad spread within the main sequence, including stars to the left of the main sequence, may suggest the presence or combination of binary stars (this investigation does not account for stellar binarity), heterogeneous foreground extinction, and photometric errors. 

The juxtaposition of observed photometry and theoretical isochrones approximates ages within our stellar population and offers evidence for a significant star-forming event spanning $\sim$25 to 40 Myrs ago, as well as consistent star formation within the SMC Wing throughout the past $\sim$100 Myr. These findings imply sustained, active star formation throughout an extended past, despite apparently low gas and dust densities within the SMC Wing, which is consistent with \citet{1964MNRAS.127..429W} and \citet{2015MNRAS.449..639R}.

\begin{figure*}[!ht]
  \includegraphics[width=\textwidth]{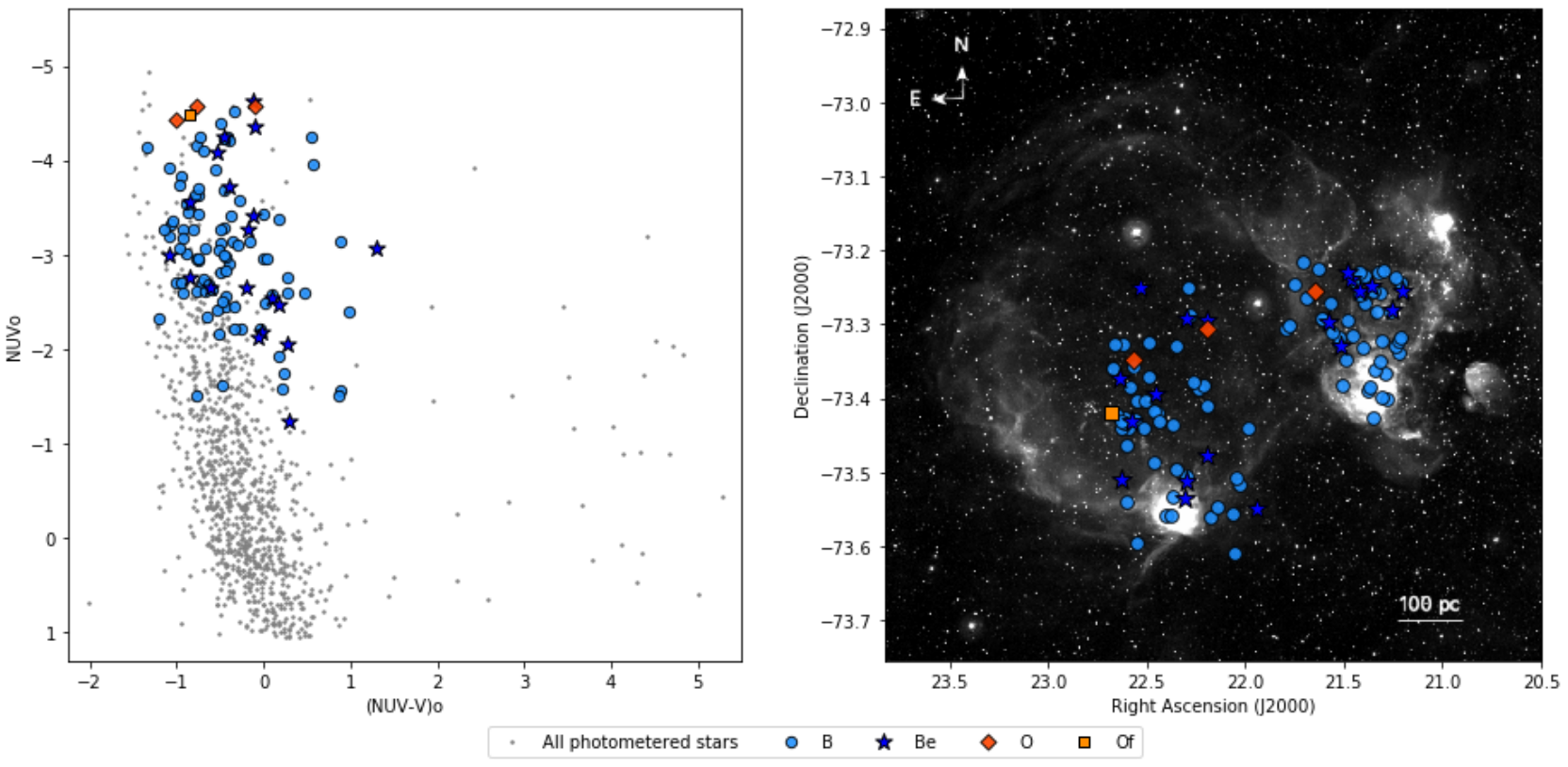}
    \caption{(a) Gray: NUV$_\circ$ vs. (NUV-V)$_\circ$ CMD for our photometered stars. Blue, Orange: Stars within the combined photometric and spectroscopic stellar sample marked by their stellar spectral classification. (b) MCELS narrow-band H$\alpha$ image of the SMC Wing highlighting the supergiant shell SMC-SGS~1. Points indicate the locations of photometered stars with marker colors and shapes indicating stellar classification.}
    \label{fig:photometric_spectroscopic_spatial}
\end{figure*}

\textit{Stellar classifications and exceptionally bright, blue stars.} 
Our combined photometric and spectroscopic sample contains 3 O stars, 1 Of star, 91 B stars, and 19 Be stars (Figure~\ref{fig:photometric_spectroscopic_spatial}). The collection of very bright, blue stars within the upper-left-hand corner of Figure~\ref{fig:cmd_isochrones}, as well as the presence of H~{\sc ii} in the SMC Wing, suggests that star formation continues into the present. Spectroscopic observations confirm the presence of both young and evolved OB stars encompassed by SMC-SGS~1. Simultaneously, these objects potentially expose gaseous heterogeneity within the supergiant shell SMC-SGS~1. To correct for local extinction within the SMC Wing, we adopted a segmented foreground screen correction for NUV$_\circ$ and V-band observations (Section~\ref{sec:observations}, Table~\ref{tab:photometric_corrections}). This correction inherently assumes a uniform gaseous structure for SMC-SGS~1 and does not account for local over- and under-densities within the shell. Therefore, while these exceptionally bright, blue stars are likely inherently bright, they perhaps additionally occupy relatively low-density pockets within SMC-SGS~1, leading to an overestimate of their foreground extinction.

\textit{Runaway stars:} 
\citet{2018ApJ...867L...8O} and \citet{2019AA...625A.104R} investigate the potential presence of runaway OB stars encompassed by SMC-SGS~1. Given the long duration of star formation and the likely number of supernovae in the SMC Wing, they expect the time between SNe~{\sc ii} events to be shorter than massive stellar lifetimes. Therefore, such SNe~{\sc ii} events provide the dynamical impetus for a predicted population of runaway OB stars within the investigated region. \citet{2019AA...625A.104R} list 25 potential runaway stars with unusually high or low radial velocities (\citet{2019AA...625A.104R} Figures~9,10), and \citet{2018ApJ...867L...8O} use proper motions from Gaia to search for possible runaways among optically bright stars. Figure~\ref{fig:velocities}b illustrates the stars in our combined photometric and spectroscopic sample that would be included in the \citet{2019AA...625A.104R} analysis of potential runaway stars. Within this figure, marker size signifies the difference between each star's radial velocity and the population median radial velocity. Due to the particular clustering of stars with relatively low radial velocities, we suspect that they may represent a separate system or OB association, as well as potential runaways.

\begin{table}[!ht]
  \centering
  \begin{threeparttable}
  \caption{Observed and theoretical mass calculations}
  \label{tab:mass_calculation}
  \begin{tabular}{p{6.58cm}r}
    \hline
\noalign{\smallskip}
Mass calculation element & Value  \\
        \hline
        \noalign{\smallskip}
        \textbf{Supporting values:} &  \\
        \noalign{\smallskip}
        IMF (1) & -2.4\\
        \noalign{\smallskip}
        Observed stellar mass range, based on isochrone log(age/year) = 7.4 [M$_\odot$] & [3, 10]\\
        \noalign{\smallskip}
        Theoretical stellar mass range, upper bound determined by (2) [M$_\odot$] & [0.1, 43]\\
        \noalign{\smallskip}
        Ratio of theoretical to observed stellar mass & 5.4\\
        \noalign{\smallskip}
        \noalign{\smallskip}
        \textbf{Mass calculation:} &  \\
        \noalign{\smallskip}
        Observed stellar mass for photometered stars, based on match to isochrone models [M$_\odot$] & 5400 \\
        \noalign{\smallskip}
        Theoretical stellar mass for the investigated region [M$_\odot$] & 30000 \\
        \hline
        \noalign{\smallskip}
        \end{tabular}
                \begin{tablenotes}
        \small
        \item References. (1) \citep{2001ApJ...554.1274C}; (2) \citep{2019AA...625A.104R}.
        \end{tablenotes}
        \end{threeparttable}
\end{table}

\subsection{The relationship between the stellar population and gaseous environment, SMC-SGS~1}

\textit{Gaseous and stellar radial velocities.} 
H$\alpha$ observations that were taken radially across SMC-SGS~1 reveal a heliocentric radial velocity of $\sim$165 km~s$^{-1}$ $\pm$ 5~km~s$^{-1}$ along the length of the slit. Observations of neutral gas within SMC-SGS~1 (Figure~\ref{fig:velocities}, red) illustrate peak heliocentric H\,{\sc i} radial velocities at $\sim$165 km~s$^{-1}$ and $\sim$180 km~s$^{-1}$, with a local minimum between those peaks at $\sim$170 km~s$^{-1}$. We interpret these observations as the leading, trailing and central components of the shell, respectively. Stellar radial velocities demonstrate a significant peak at $\sim$ 170 km~s$^{-1}$ for our combined photometric and spectroscopic sample. The alignment of the H\,{\sc i} local minimum and stellar global maximum at this velocity suggests a direct relationship between local star formation and the gaseous supergiant shell SMC-SGS~1.

\textit{Gaseous and stellar metallicities.} 
Congruence between stellar and gaseous metallicities within our investigated region further implies an interconnection. The SMC-SGS~1 H$\alpha$ emission line was clearly detected and had an intensity of about four times that of the night sky H$\alpha$ emission line. However, the width of this line was not resolved because it was blended with a faint night sky line. Therefore, we did not resolve the internal motions of the expanding shell either by finding velocity gradients or by detecting a broadened H$\alpha$ emission line. These results are consistent with slow motions for ionized gas and a generally quiescent gaseous environment. Our spectra include faint [N\,{\sc ii}] emission with a flux of $\sim$ 5\% of the H$\alpha$ line, as well as stronger emission lines from the [S\,{\sc ii}] doublet. Both of these line ratios are typical of H\,{\sc ii} regions in the low-metallicity SMC (e.g., \citet{1976ApJ...203..581P}, \citet{1977ApJ...216..706D}), and they are consistent with low metallicities in massive stars, which was inferred by \citet{2019AA...625A.104R}.

\textit{Timeline for shell expansion.} 
We observe an alignment between the SMC-SGS~1 expansion time and the ages of our stellar population, offering temporal evidence for their mutual formation. To find an expansion age for the shell, we take the HI local minimum at $\sim$170 km~s$^{-1}$ (Figure~\ref{fig:velocities}a, red) as the shell center, and the neighboring peaks at $\sim$160 km~s$^{-1}$ and $\sim$180 km~s$^{-1}$ as its leading and trailing edges, respectively. Adopting the observed radius of $\sim$300~pc \citet{1980MNRAS.192..365M}, we calculate an expansion velocity of 10 km~s$^{-1}$ for the supergiant shell. The standard model for shell expansion offers a lower bound on the expansion age of $t_{exp} \approx 0.6 * (R_{shell}\ /\ V_{exp}) \approx$ 20~Myr. However, the shell shows no evidence of having originated from a single star cluster, suggesting that it does not follow the standard model. SMC-SGS~1 more closely resembles a picture of supergiant shell production by noncoeval (i.e., ``not of the same age'') star formation within an OB association \citep{1995ApJ...444..663S}. In this scenario, potential ages range $\sim$ 20 - 40 Myr for large shells with $V_{exp} \sim$ 10~km~s$^{-1}$ (see \citet{1995ApJ...444..663S} Figure~4), with exact age depending on the detailed history of the region. The presence of cold gas clouds \citep{2008PASP..120..972N} may represent an additional source of radiative cooling within SMC-SGS~1. Such clouds would diminish the interior pressure and slow shell expansion as compared to predictions of standard adiabatic models (i.e., ``poisoning'' the shell, \citet{1995ApJ...444..663S}). Given the above considerations, the estimated shell expansion ages are consistent with the timescale of the most recent epoch of star formation near SMC-SGS~1.

\textit{Structure of SMC-SGS~1.} 
The physical stability of SMC-SGS~1 provides insight into the SMC Wing environment. SMC-SGS~1 maintains a remarkably well-defined structure despite its size, suggesting that both the shell and the local ISM have relatively low gas densities. \citet{2019AA...625A.104R} find weak local stellar winds throughout the SMC Wing, concluding that supernovae contribute the large majority of mechanical energy necessary to sustain the shell's structure. We estimate an H\,{\sc i} mass of $\sim$10$^6$~M$_{\odot}$ for our investigated region \citep{1999MNRAS.302..417S}. With the calculated expansion velocity of 10~km~s$^{-1}$, the supergiant shell would require $\sim$100 supernovae to support its observed momentum \citep{2019AA...625A.104R}. Given its young stellar population and the age of its most recent star-forming event, this region could reasonably produce the number of supernovae necessary to form the supergiant shell.

\begin{figure}[!ht]
	\includegraphics[width=1.02\columnwidth]{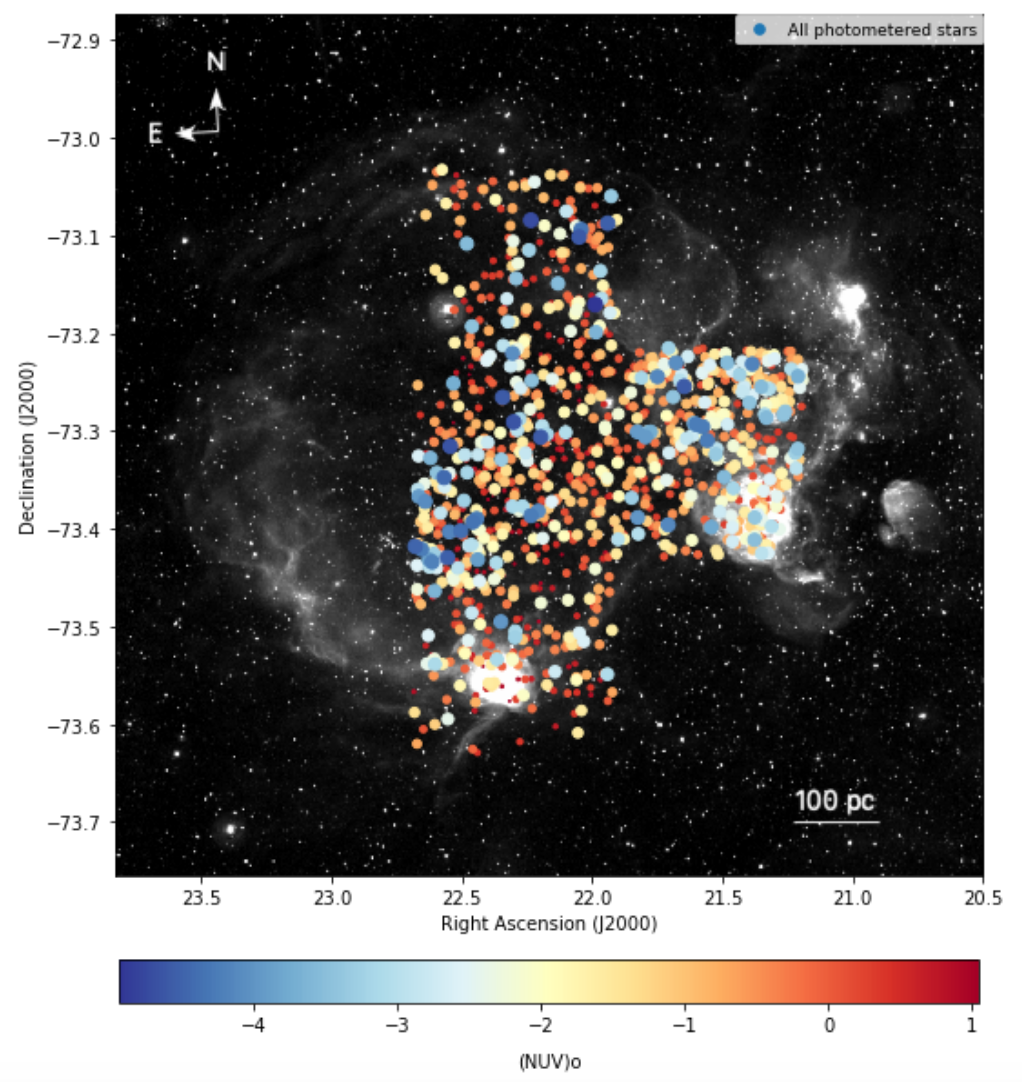}
    \caption{MCELS narrow-band H$\alpha$ image of the SMC Wing highlighting the supergiant shell SMC-SGS~1. Points indicate the locations of photometered stars, with marker sizes and colors indicating stellar (NUV)$_\circ$ magnitude.}
    \label{fig:photometric_spatial}
\end{figure}

\textit{Spatial distribution.} 
The spatial distribution of bright stars within our investigated region suggests a combination of stimulated star formation due to the expansion of SMC-SGS~1 and a stochastic mode of star formation that is independent of shell expansion. Given the prominence of SMC-SGS~1, one may expect star formation to propagate systematically at the shell edge as it expands into the surrounding environment. The location of the young star cluster NGC~602 on the southern edge of SMC-SGC~1 supports this narrative, and we infer that shell expansion has some influence on local star formation. However, this propagation would further result in a stellar age and luminosity gradient along the radial axis of SMC-SGS~1, with relatively faint stars within its center and relatively bright stars at its outermost edge. Within our investigated region, Figure~\ref{fig:photometric_spatial} reveals no such gradient; we observe an equal distribution of bright, young stars on the edge of the shell as within its volume. {\tt DAOPHOT} selection criteria that under-resolve stars in high-density clusters, combined with high gas concentrations at the southern and western edges of the shell (Figure~\ref{fig:smc_investigated_region}b) may explain the dearth of bright photometered stars at these locations. Massive stars may form in relatively long-lived clusters \citep[e.g.,][]{2018MNRAS.476..381W}, and thus some fraction of the young stars are likely located within compact systems that were not resolved by {\tt DAOPHOT}. Furthermore, we expect younger regions to have greater gas concentrations, potentially leading to higher local extinction and preferential reddening of light from young stars. Nonetheless, the observation of bright photometered stars within the volume of SMC-SGS~1 suggests some influence from a stochastic mode of star formation.

\textit{Molecular clouds.} 
The region containing SMC-SGS~1 shows little evidence for a substantial molecular ISM \citep[e.g.,][]{2008PASP..120..972N}. H\,{\sc i} pervades the SMC Wing; however, in the area surrounding SMC-SGS~1, its surface density (measured with a low-resolution map that averages over the local structures) is only about 10~M$_{\odot}$~pc$^{-2}$ (\citet{1981PASAu...4..189M}. This places SMC-SGS~1 on a boundary between regions where the ISM is dominated by H\,{\sc i} and where the molecular fraction is significant \citep[e.g.,][]{2013MNRAS.436.2747K}. The tidal structure of the SMC Wing, together with the local quiescence of the ISM \citep{2010ApJ...708.1204B} provides an environment where gravitationally-unstable molecular clouds may be slow to form \citep{2013MNRAS.436.2747K}. The extended star formation that is observed within the volume of the shell may support this slow formation timescale, with the passage of the expanding shell itself creating local instabilities. Alternatively, existing local molecular clouds may have remained stable and star-forming even as ionized gas was expelled to form SMC-SGS~1, only later reaching the point of collapse. In either scenario, a softly expanding shell that sweeps out diffuse gas but passes over high-density molecular clouds may hold significant implications for star formation and feedback in low-density environments. We suggest a further investigation of the molecular cloud distribution within the SMC Wing, particularly given strong evidence for a molecular cloud directly associated with the cluster NGC~602 \citep{2008PASP..120..972N}.

\subsection{A comparison among dwarf galaxy stellar populations}

\textit{Star formation intensity.} 
The stellar population encompassed by SMC-SGS~1 may offer an interesting comparison with star-forming regions in other low-density environments \citep[e.g.,][]{2010AJ....140.1194B,2010AJ....139..447H,2016ApJ...832...85T}. Small-scale regions within dwarf galaxies tend to be characterized by extended star-forming events, which are separated in time by considerably longer quiescent periods \citep{2008ApJ...681..244B,2012ApJ...744...44W,2016ApJ...832...85T}. We observe this same trend within SMC-SGS~1. Our investigation covers an area of 0.20~kpc$^2$ and reveals $\sim$ 3$\times~10^4$~M$_{\odot}$ of star formation over the past $\sim$ 25~Myr (Table~\ref{tab:mass_calculation}). Based on these values, the average star-formation rate and star-formation intensity are $\sim$ 1.2 $\times$ 10$^{3}$~M$_{\odot}$~yr$^{-1}$ and $\sim$ 6 $\times$ 10$^{-3}$~M$_{\odot}$~kpc$^{-2}$~yr$^{-1}$, respectively, thus falling within a regime of low-intensity star formation. This long duration of star formation, when spread over a wide spatial area,  resembles the low-intensity star-forming regions found in star-forming dwarf galaxies, as well as other regions within the SMC.

\textit{Dual mode of star formation.} 
The observed stellar population follows a scenario of combined stimulated and stochastic modes of star formation within a low-density environment \citep{2008PASP..120..972N, 2016ApJ...832...85T}. This pattern is further observed in low-mass-density dwarf irregular galaxies \citep{2000ApJ...529..201S, 2017MNRAS.464.1833E}. The combination of star-formation modes yields a stellar population dominated by B stars with relatively long lifetimes (Figure~\ref{fig:photometric_spectroscopic_spatial}, \citet{2019AA...625A.104R}), and thus may be a significant source of ultraviolet luminosity over substantial timescales. As suggested in the case of dwarf galaxies, \citep[e.g.,][]{2008ApJ...681..244B,2009ApJ...706..599L,2009ApJ...695..765M,2010AJ....139..447H,2012ApJ...744...44W,2016ApJ...817..177L}, this type of active star formation results in regions of bright ultraviolet luminosity and only modest H$\alpha$ luminosity.

\section{Conclusions}\label{sec:conclusions}

We present a photometric survey of stellar populations within the SMC Wing and their relationship to the supergiant shell SMC-SGS~1, the photometric complement to a spectroscopic study of a similar region by \citep{2019AA...625A.104R}.

We combine near-ultraviolet and optical (V-band) photometry for $\sim$1000 stars encompassed by SMC-SGS~1 and adopt spectroscopic information for $\sim$ 10\% of the photometric population. Our (NUV)$_\circ$ vs. (NUV-V)$_\circ$ color-magnitude diagram demonstrates a well-defined main sequence, relatively bright and blue stars indicating a significant population of young stars, and evolved stars of various ages. Superimposed isochrone models from the University of Padova effectively quantify stellar ages for our investigated region, from which we find many stars with ages spanning $\sim$ 25 - 40~Myr. We interpret these stars as indicative of an extended star-formation event over this timescale, consistent with the findings of \citep{2019AA...625A.104R}. Combined photometric, spectroscopic, and spatial analyses confirm the presence of OB stars and demonstrate that young stars are scattered evenly throughout the investigated region.

Our stellar population and the supergiant shell SMC-SGS~1 demonstrate agreement in radial velocity, age, and metallicity, suggesting that they comprise an interconnected physical system with a mutual origin. However, we do not observe a centralized star cluster whose formation may have produced such a dramatic structure. Instead, we observe young stars both at the edge and within the volume of SMC-SGS~1, suggesting a combination of stimulated and stochastic star-formation modes. Star formation interior to the shell implies either that molecular clouds formed after the passage of SMC-SGS~1, or that existing molecular clouds survived the expansion of the shell, remaining stable and star-forming event as local gas and dust were swept out.

Adopting a standard initial mass function, we calculate a lower bound on the young stellar mass of $\sim 3 \times$10$^4$~M$_{\odot}$ for our investigated region, corresponding to a star-formation intensity of $\sim$ 6 $\times$ 10$^{-3}$~M$_{\odot}$~kpc$^{-2}$~yr$^{-1}$. This star-formation intensity places SMC-SGS~1 within the \citet{2000ApJ...530..277E} ``star formation in a crossing time'' correlation (i.e., a standard pattern for the duration of star-formation events in low-density, tidally-disturbed regions). Given its location near the tip of the SMC Wing, our investigated region offers a local example of active star formation within a tidally-induced, dynamically-calm environment \citep[e.g.,][]{2003MNRAS.339.1135Y, 2014MNRAS.442.1663D}. Therefore, it shares many properties with instances of low-intensity star formation in the outer regions of galaxies \citep[see also \S4.3][]{2007ApJS..173..538T,2013MNRAS.436.2747K, 2017ASSL..434..115E}.

\begin{acknowledgements}
	The authors thank Sne\v{z}ana  Stanimirovi\'{c} and McClure-Griffths for supplying the H\,{\sc i} imaging data, as well as Jochen M. Braun for making their thesis photometry publicly available; these archival data were essential for this project. We thank Ralf Kotulla for creating the mosaic of GALEX images that was fundamental and invaluable for our photometry. Some of the observations reported in this paper were obtained with the Southern African Large Telescope (SALT) as part of program 2016-2-SCI044, and thus we thank the SALT observatory team for their key role in obtaining our spectroscopic data. The authors thank Ellen Zweibel for her informative discussion of the astrophysics of supergiant shells. Part of this work was completed as Leah M. Fulmer's undergraduate senior thesis, including support from the University of Wisconsin-Madison and the Wisconsin Space Grant Consortium. Leah M. Fulmer appreciates the opportunity to work at the National Optical Astronomy Observatory (NOAO) and the NOAO Data Lab during the time that this work was in progress. The authors thank the cleaning and maintenance staff of all research facilities, without whose labor this work could not be conducted. We humbly acknowledge the native lands of the Oceti Sakowin, Miami, Ho-Chunk, Sauk, Meskwaki, Peoria, Hohokam, Tohono O'odham, Puget Sound Salish, and Duwamish Tribes, where we are grateful to live and work.
\end{acknowledgements}

\bibliographystyle{aa}
\bibliography{main.bib}

\end{document}